\documentclass[aps,reprint,amsmath,amssymb]{revtex4-2}
\usepackage{graphicx}
\usepackage{dcolumn}
\usepackage{bm}
\usepackage{siunitx}
\usepackage{lipsum}
\usepackage{orcidlink}
\usepackage{titlesec}

\begin{document}

\title{Doping-induced Polyamorphic Transitions in Fluorite Oxides}
\author{Hao Yang$^{1,\orcidlink{0009-0008-6526-4644}}$}
\author{Qiaotong Luan$^2$}
\author{Qing Zhang$^{3,\orcidlink{0000-0001-8274-4270}}$}
\author{Yuhao Yue$^2$}
\author{Yawen Xu$^2$}
\author{Xiaohui Liu$^{3,\orcidlink{0000-0001-9450-8320},*}$}
\author{Zheng Wen$^{2,\orcidlink{0000-0001-5026-6585},*}$}
\author{Zhaoru Sun$^{1,\orcidlink{0000-0002-2092-9779},}$}

\email{Corresponding author. Email: liuxiaohui@sdu.edu.cn (X. L.), zwen@qdu.edu.cn (Z. W.), sunzhr@shanghaitech.edu.cn (Z. S.)}

\affiliation{
$^1$School of Physical Science and Technology, ShanghaiTech University, Shanghai 201210, China\\
$^2$College of Electronics and Information \& Shandong Key Laboratory of Micro-nano Packaging and System Integration, Qingdao University, Qingdao 266071, China\\
$^3$School of Physics, Shandong University, Ji'nan 250100, China
}

\date{\today}

\begin{abstract}
Fluorite oxides such as HfO$_2$ exhibit rich and tunable phase behavior, making them promising candidates for next generation electronic devices. A key challenge is to design amorphous HfO$_2$-based high-$k$ materials with both structural and performance stability. Here, using molecular dynamics simulations supported by experimental measurements, we reveal that Ba doping stimulates a polyamorphic transition in HfO$_2$, yielding a semi-ordered amorphous (SA) phase characterized by disordered oxygens embedded within an ordered metal sublattice. We find that this phase arises from degenerate short-range symmetry breaking modes, consistent with Pauling's parsimony rule. Notably, the SA structure is thermodynamically stable and displays a wider bandgap and higher dielectric constant than conventional random-packing amorphous structure, owing to suppressed subgap states and increased Born effective charges. We further demonstrate that this structural motif generalizes to Ba-, Sr-, and Ca-doped HfO$_2$ and ZrO$_2$, establishing a broadly applicable strategy for designing high-performance amorphous dielectrics.
\end{abstract}

\maketitle

\section{Introduction}

In complementary metal-oxide semiconductor (CMOS) technology, the dielectric properties of gate oxides are crucial for device performance and reliability \cite{RN175,RN176,RN114}. As the lateral dimension of the devices scales down, SiO$_2$ gate dielectrics ($k\mathrel{\sim}3.9$) fail due to tunneling leakage current and reliability issues \cite{RN120,RN188}. This has driven the pursuit of high-$k$ materials, which can be grown as thicker films while maintaining required capacitance \cite{RN123,RN124,RN126}. Beyond electrical performance, amorphous phases are preferred for their uniform capacitance density and stable interfaces with electrodes and channels \cite{RN127,RN128,RN129}.

Among various high-$k$ candidates, fluorite oxides, especially HfO$_2$, stand out for their superior dielectric properties and diverse structural polymorphs \cite{RN120,RN51}. Despite extensive studies \cite{RN152,RN153,RN154} demonstrating excellent performance in crystalline HfO$_2$, it remains inadequate for amorphous integration in industrial applications \cite{RN79}. Notably, HfO$_2$ is a poor glass former \cite{RN120}. Even with rapid quenching, it tends to recrystallize during subsequent annealing \cite{RN151}. Thus, developing a universal route to produce stable and amorphous high-$k$ HfO$_2$-based oxides remains a central challenge.

Contrary to conventional rapid-quench approaches, doping-based strategies have enabled amorphization under equilibrium-like conditions \cite{RN79,RN55}. However, a trade-off persists between amorphous phase stability and dielectric properties \cite{RN144,RN146,RN172,RN73,RN171,RN155,RN167,RN162}. For instance, Al doping improves amorphous stability but reduces the dielectric constant \cite{RN167}, while Y doping enhances permittivity yet promotes cubic-phase crystallization \cite{RN162}. Recent advances \cite{RN55,RN178} highlight the impact of disorder engineering on dielectric properties, such as enhanced dielectric breakdown strength exceeding the up-limit regulated by the permittivity in amorphous Hf-based oxides \cite{RN55} and ultralow-$k$ behavior in amorphous boron nitride \cite{RN178}. Nevertheless, the microscopic mechanisms linking atomic structure to dielectric response remain largely unresolved, limiting the transferability of such strategies. A key challenge lies in the limited resolution of experimental techniques, which often confirm amorphization qualitatively but fail to resolve atomic-scale structure \cite{RN179,RN186,RN187}. In this context, molecular dynamics (MD) simulations provide essential atomistic insights into structural-property relationships in amorphous dielectrics.

In this work, taking Ba-doped HfO$_2$ as a model system, we combine MD and first-principle simulations with experimental measurements to investigate structural change induced by substitutional doping, and identify polyamorphic transitions across all alkaline-earth dopants. Our simulations reveal a semi-ordered amorphous (SA) structure with proper doping concentration, characterized by disordered oxygens within an ordered metal sublattice formed by Ba and Hf, consistent with extended X-ray absorption fine structure (EXAFS) measurements. We show that this SA phase results from a degeneracy of short-range symmetry breaking modes, in accordance with Pauling's parsimony rule. More importantly, the SA phase demonstrates enhanced thermodynamic stability and dielectric performance, i.e., higher permittivity, and superior insulation, compared to conventional random-packing amorphous (RA) structures. Furthermore, we demonstrate the generality of this structural motif across alkaline-earth-metal-doped HfO$_2$ and ZrO$_2$ systems, highlighting its robustness and compositional tunability.

\section{Methods}
Classical MD simulations were performed to investigate the doping-induced structural transition, using Ba-doped HfO$_2$ (BHO) as a representative system. 
MD simulations were carried out using the Large Atomic/Molecular Massively Parallel Simulator (LAMMPS) package \cite{RN9}. The initial configuration consisted of a fluorite-structured HfO$_2$ system with 768 atoms ($N_{\text{Hf}} = 256$, $N_{\text{O}} = 512$), where Hf atoms were partially substituted with alkaline earth metals (Ba, Sr, Ca) at varying concentrations (3\%, 9\%, 12\%, 19\%, and 37\%).Charge neutrality was maintained by randomly removing an equivalent number of oxygen atoms. To account for polarization effects, the core-shell model was applied \cite{RN25}, and interatomicinteractions were described using the Buckingham potential \cite{RN2}. The system was equilibrated at 1500 K, then gradually annealed and equilibrated to 873 K and subsequently to 300 K, mimickingthe experimental synthesis process. Electronic properties calculations were performed using the Vienna \textit{ab-initio} Simulation Package (VASP) \cite{RN12} based on density functional theory (DFT). The Perdew-Burke-Ernzerhof (PBE) exchange-correlation functional was utilized \cite{RN13}, while the density of states (DOS) was obtained using the hybrid PBE0 functional to accurately capture the bandgap \cite{RN135,RN105,RN106}. Structural characterization of the amorphous films was conducted via X-ray diffraction (XRD) and extended X-ray absorption fine structure (EXAFS) measurements. Further details on modeling, computational parameters, and experimental procedures are provided in the Supplementary Information (SI). 

\section{Results}

\begin{figure}[b]
\centering
\includegraphics[width=0.85\columnwidth]{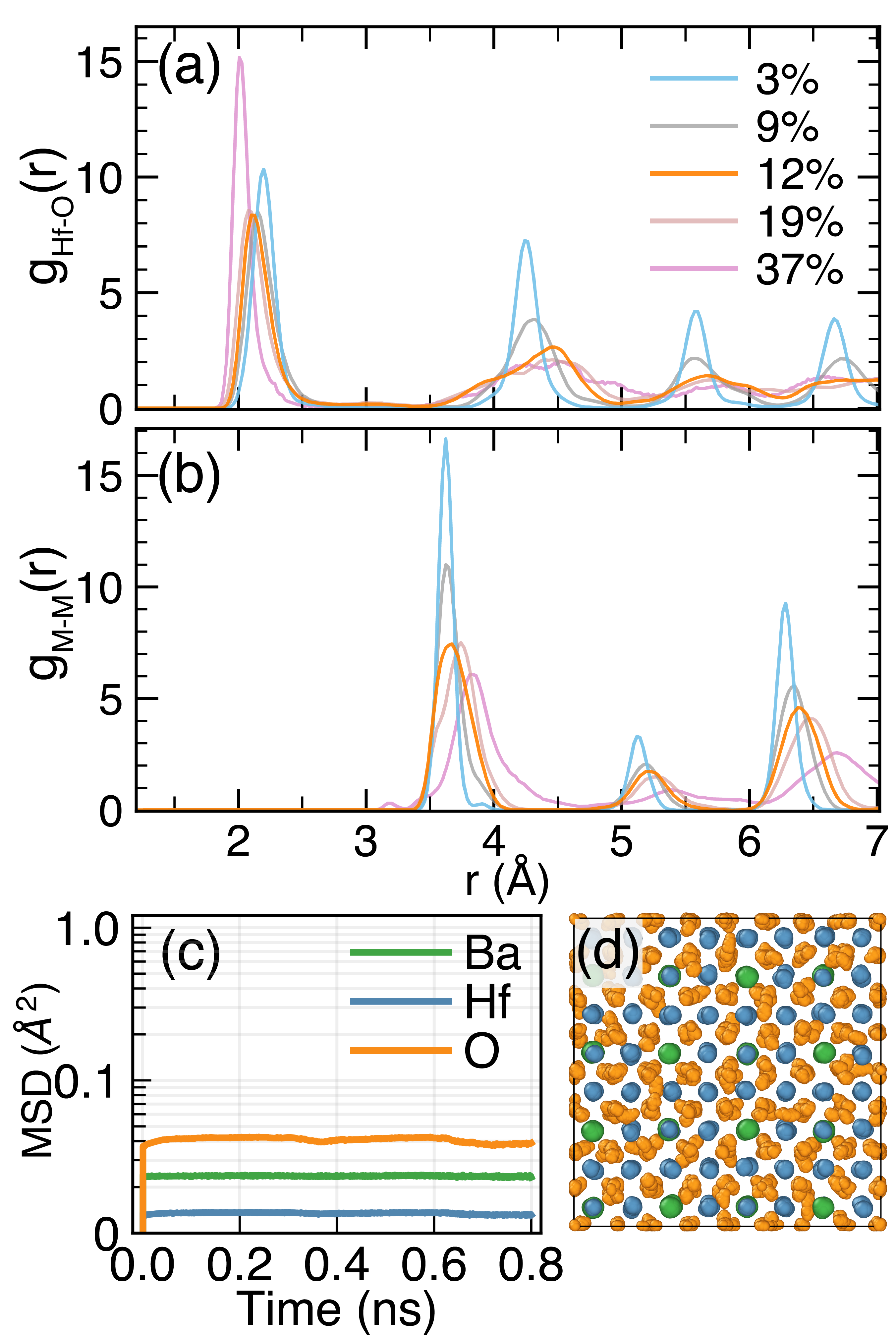}
\caption{
(a) Hf-O and (b) metal-metal (M-M, M = Ba or Hf) radial distribution functions (RDFs) of Ba-doped HfO$_2$ with doping concentrations ranging from 3\% to 37\% (corresponding to BHOx, x=3-37) at 300 K. (c) mean squared displacement (MSD) and (d) atomic trajectory overlaps of atoms in BHO12 at 300 K.
}
\label{fig:rdf}
\end{figure}

We characterize Ba-doping-induced structural changes in HfO$_2$ across 3\% to 37\% doping (BHO3 - BHO37) using Hf-O and metal-metal (Ba/Hf) radial distribution functions (RDFs), which describes the probability of finding neighboring atoms at a given distance from a reference atom \cite{RN198}. Crystalline phases show sharp RDF peaks reflecting long-range order, while amorphization leads to peak broadening and intensity decay toward $g(r)=1.0$, consistent with uniform bulk density. As shown in Figs.~\ref{fig:rdf}(a) and \ref{fig:rdf}(b), peak intensities in both $g_{\text{Hf-O}}(r)$ and $g_{\text{M-M}}(r)$ decrease with increasing Ba content up to 19\%, indicating progressive structural distortion and increased amorphization. Notably, the amorphous structure emerging at 12\% - 19\% doping differs significantly from that of a conventional, random-packing amorphous (RA) phase. In the intermediate range (4-8 \AA), $g_{\text{Hf-O}}(r)$ exhibits a progressive loss of peak features, converging toward 1.0, while $g_{\text{M-M}}(r)$ retains well-defined peaks separated by near-zero valleys. This contrast indicates that the metal sublattice remains ordered, whereas the oxygen sublattice becomes disordered. Here, we denote this configuration as a semi-ordered amorphous (SA) structure.

The structural differences between SA and RA systems are also supported by experiment. To validate this, we compare RDFs of BHO12 in SA and RA forms with extended X-ray absorption fine structure (EXAFS) data. The RA structure is generated by melting and quenching the SA structure (see SI Sec.~M1). The experimentally measured Hf-O coordination number (CN$_{\text{Hf-O}}$) lies between 6.75 and 6.85, closely matching that of SA BHO12 (6.81) but deviating significantly from the RA value (5.99) (Table S1). Additionally, EXAFS peak positions are more consistent with the SA structure and differ from those of the RA structure (Fig.~S1). These results, supported by both simulations and experiments, confirm that Ba doping induces a semi-ordered amorphous phase characterized by disordered oxygen atoms embedded within an ordered metal framework.

\begin{figure*}[t]
\includegraphics [width=0.9\textwidth]{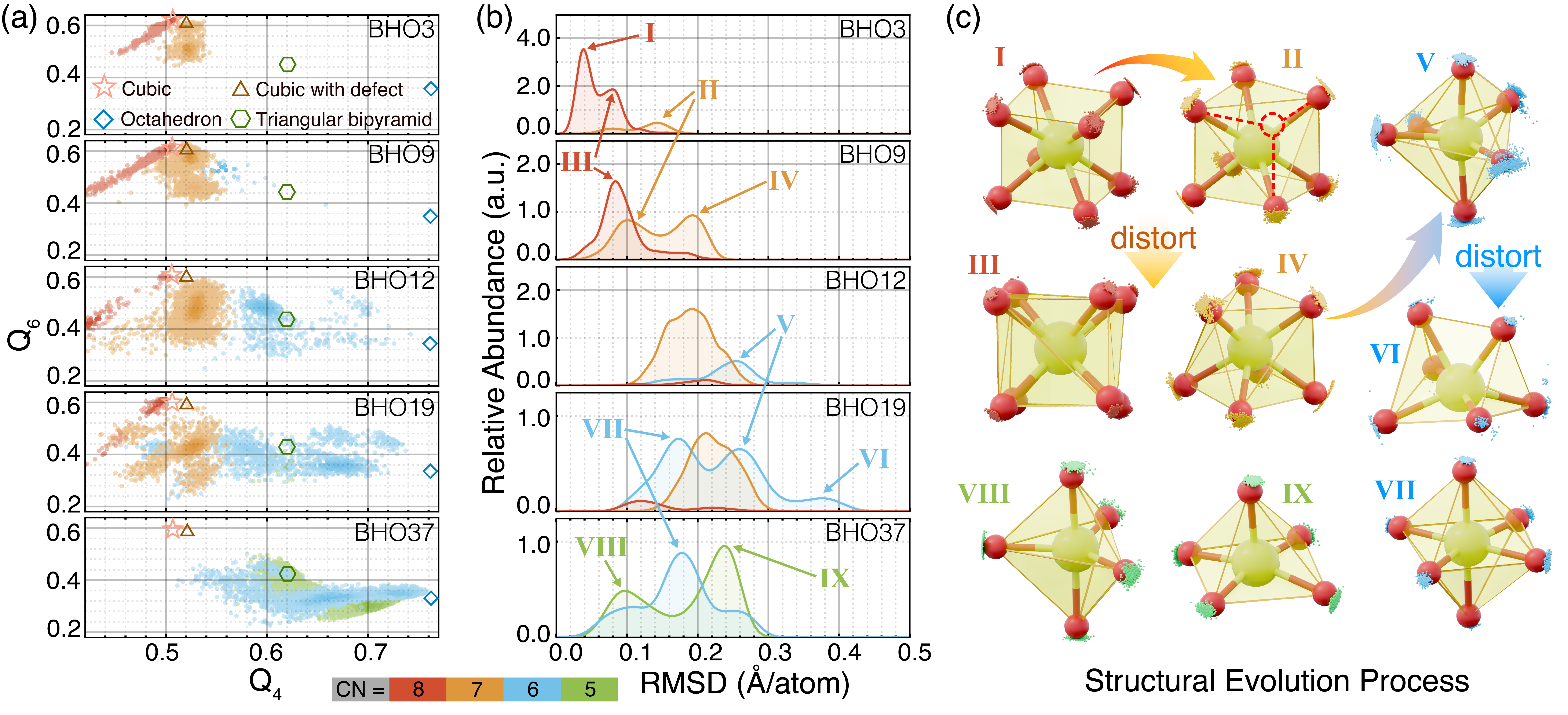}
\caption{(a) Bond-orientational order development for cages depicted by Q$_4$-Q$_6$, with reference points marking reference cage types: cubic (0.51, 0.63), cubic with defect (0.52, 0.63), octahedron (0.76, 0.35), and triangular bipyramid (0.62, 0.45). (b) Relative abundances of different cage types. The area sizes are normalized to facilitate direct comparisons across doping concentrations. Color mapping in (a) and (b) reflects different coordination numbers of Hf-O (CN$_{\text{Hf-O}}$) (c) Representative cages illustrate structural evolution process with increased doping concentration. Dots indicating cage positions after rotation for maximal superimposition.}
\label{fig:cages}
\end{figure*}

We further illustrate the SA structure by analyzing the mean squared displacement (MSD) and atomic trajectories, as shown in Figs.~\ref{fig:rdf}(c) and \ref{fig:rdf}(d). At 300 K, Ba and Hf atoms exhibit an ordered lattice with minimal diffusivity ($<6.25\times10^{-6}$~\AA$^2$/ps), consistent with thermal vibrations \cite{RN180,RN181}. Oxygen atoms, though disordered, exhibit similarly low mobility, confirming the stability of the SA structure at ambient conditions. Upon experimental annealing temperature of 873 K (Fig.~S2), oxygens undergo significant lattice hopping, as indicated by continuous site-to-site trajectories with orders larger diffusivity ($\sim1.25\times10^{-3}$~\AA$^2$/ps). In contrast, metal atoms remain localized, with MSDs below $7.5\times10^{-5}$~\AA$^2$/ps. These results demonstrate that the SA phase exhibits a unique structural motif: oxygen disorder confined within an ordered metal framework, which degrades at elevated temperatures due to thermally activated oxygen diffusion.

Building on the structural feature of the SA phase, we investigate its formation mechanism. RDFs reveal a systematic Hf-O bond contraction from 2.11 \AA\ (HfO$_2$-like) to 2.01 \AA\ (BaHfO$_3$-like), linking the two structural end members. Since the key difference lies in Hf-O cage symmetry, we analyze local symmetry changes using bond-orientational order parameter (BOP) analysis. Based on spherical harmonics of degree $l$, BOP is widely utilized to quantify local symmetries in disordered systems \cite{RN182}. We focus on $l$ = 4 and 6 (denoted Q$_4$ and Q$_6$), which effectively distinguish cubic and octahedral cages \cite{RN197}, corresponding to basic Hf-O units in fluorite and perovskite structures (see SI Sec.~M2). Figure 2(a) maps Hf-O cages onto the Q$_4$-Q$_6$ plane for doping concentrations from 3\% to 37\%, with high-symmetry references from the Thomson problem and Schottky-defect model \cite{RN45,RN183}. As Ba doping increases, the Hf-O coordination number decreases from eight- to five-fold, accompanied by a reorientation of local bonding environments. In the SA phase regime (12\% - 19\%), the cage distribution shifts toward lower Q$_4$ and Q$_6$, indicating reduced local symmetry. Outside this range, clusters align toward cubic (pre-SA) and octahedral (post-SA) values, corresponding to fluorite- and perovskite-like order.

We identify representative Hf-O cage geometries tracing the transformation from cubic- to octahedral-like cages and observe a process of symmetry breaking followed by symmetry restoration. As shown in Figs.~\ref{fig:cages}(b) and \ref{fig:cages}(c), nine cage types are identified via root-mean-square deviation (RMSD) analysis \cite{RN18}. Dots in Fig.~\ref{fig:cages}(c) represent bond orientations determined by the Kabsch algorithm \cite{RN118}, which minimizes RMSD to optimize structural alignment (see SI Sec.~M2). As Ba concentration increases, the system transitions from fluorite-like cages (types I - III) exhibiting cubic symmetry, to distorted intermediates (types IV and V) reflecting pronounced symmetry breaking. At higher doping levels, these distorted cages give way to type VII cages with octahedral-like symmetry, indicative of a perovskite-like phase. Concurrently, fivefold-coordinated cages (types VIII and IX) emerge as structural intermediates, bridging the transition between cubic and octahedral motifs. This progression illustrates a continuous structural transformation driven by successive symmetry breaking and restoration.

Notably, the SA phase, exemplified by 12\% doping, emerges through a distinct polyamorphic transition driven by unconventional short-range symmetry breaking. On one hand, the cage in SA phase shares similar geometries with those in fluorite-like or perovskite-like systems but with different abundances. As shown in Fig.~\ref{fig:cages}(b), the SA phase is dominated ($>90\%$) by type IV cages with minor contributions from types II and V, although precise quantification is limited by broad RMSD distributions (0.10-0.28~\AA/atom). These distorted cages appear sparsely in fluorite- or perovskite-like phases but are highly prevalent in the SA phase, indicating maximal short-range symmetry breaking. On the other hand, the symmetry breaking mechanism in the SA phase differs fundamentally from that in the RA phase. The structural similarity of seven-coordination cages in SA indicates degeneracy of symmetry breaking modes. In contrast, the RA phase (Fig.~S4) exhibits a broad mixture of cage types (six-fold: 67\%, seven-fold: 17\%, eight-fold: 16\%) with wider RMSD spread (0.1-0.5~\AA/atom), reflecting non-degenerate, more disordered symmetry breaking. Moreover, this structural contrast governs the metal framework arrangements. Consistent with Pauling's fifth rule of parsimony \cite{RN196} which favors crystals with few distinct polyhedral types, the SA phase exhibits degenerate cage geometries that enable crystalline-like ordering in the metal sublattice. 

Overall, these findings elucidate a doping-induced transition of short-range symmetry. The SA phase emerges from short-range symmetry breaking, yet its degenerate distortion modes distinguish it from the RA phase, reflecting a form of polyamorphism. Aditional simulations under varied initial conditions (Fig.~S5) confirm the robustness of these trends.

\begin{figure}[t]
\centering
\includegraphics[width=0.9\columnwidth]{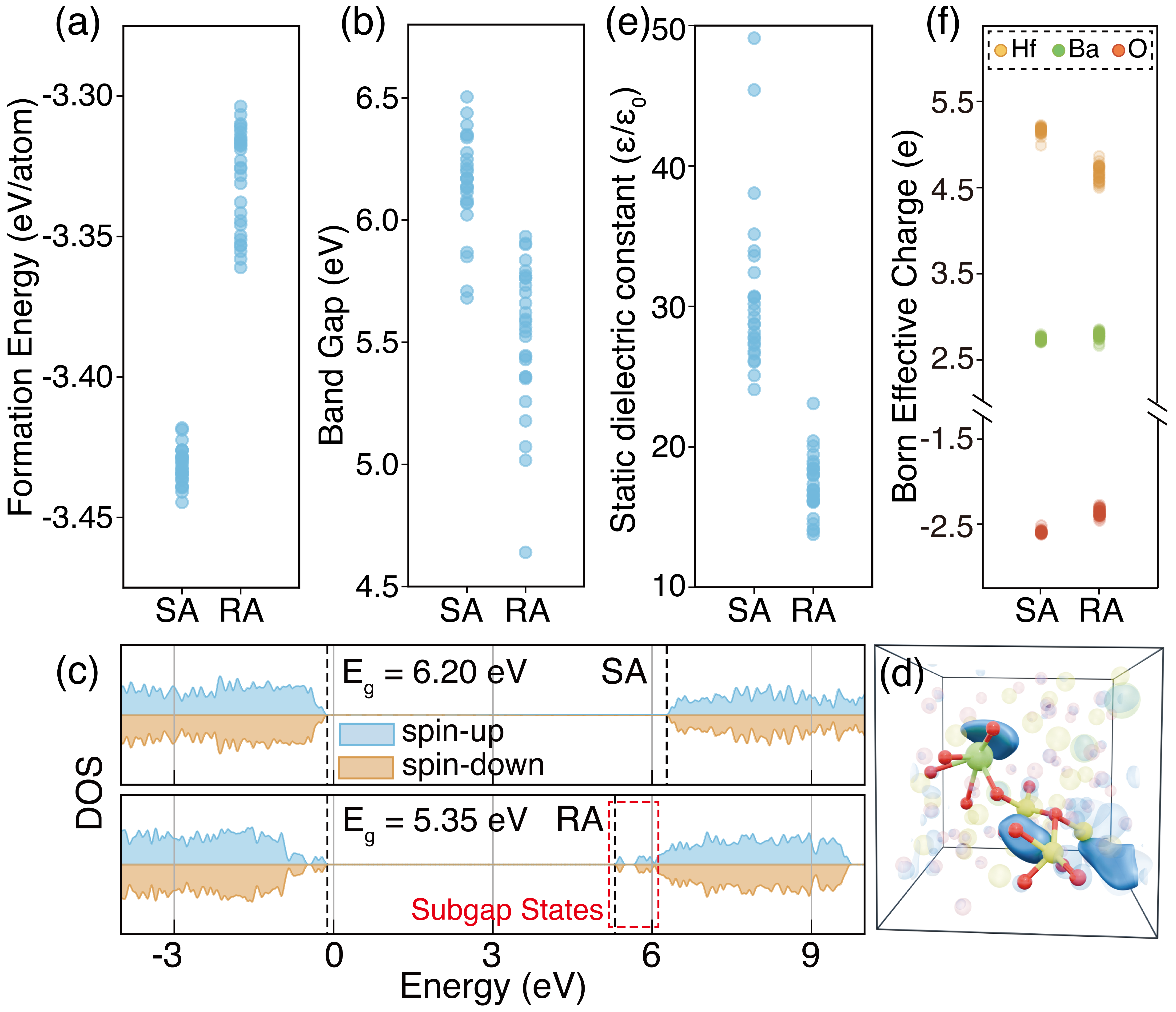}
\caption{
Structural and electronic properties of BHO12 in the SA and RA phases: (a) formation energy, (b) bandgap, and (c) density of states (DOS), with the Fermi level set to zero and black dashed lines indicating the valence band maximum (VBM) and conduction band minimum (CBM). (d) Charge density of CBM state in the RA phase. (e) Static dielectric constant and (f) Born effective charge.
}
\label{fig:properties}
\end{figure}

Based on our comprehensive understanding of the doping-mediated SA structure, we focus on its feasibility as a gate oxide in comparison to the conventional RA phase, and reveal the critical role of the ordered metal sublattice. To ensure statistical reliability, we randomly selected two sets of 25 snapshots from the SA and RA trajectories of BHO12 (see Sec.~2 in SI). First, formation energy calculations confirm that the SA phase is thermodynamically more stable than the RA phase, consistent with experimental observations. The SA phase exhibits a lower average formation energy (-3.44 eV/atom) than the RA phase (-3.33 eV/atom), along with a narrower energy distribution due to the preserved metal sublattice order (Fig.~\ref{fig:properties}(a)). Furthermore, the SA phase displays enhanced insulating behavior. As shown in Figs.~\ref{fig:properties}(b) and \ref{fig:properties}(d), the RA phase exhibits smaller bandgap due to the suppression of subgap states near the conduction band minimum (CBM), which are prevalent in the RA phase due to metal sublattice disorder, including dangling bonds and direct metal-metal interactions. Finally, the SA phase exhibits superior dielectric performance. As shown in Figs.~\ref{fig:properties}(e) and \ref{fig:properties}(f), it has a higher static dielectric constant than the RA phase, in agreement with prior experimental trends \cite{RN55}. This enhancement originates from the ordered metal sublattice, which allows Hf atoms to gain charge from a greater number of surrounding oxygen atoms, increasing the Born effective charge and thus the dielectric response. Taken together, these results demonstrate that the ordered metal sublattice enables the SA phase to overcome the conventional trade-off between bandgap and dielectric constant, positioning it as a compelling candidate for next-generation gate dielectric materials.

\begin{figure}[b]
\centering
\includegraphics[width=\columnwidth]{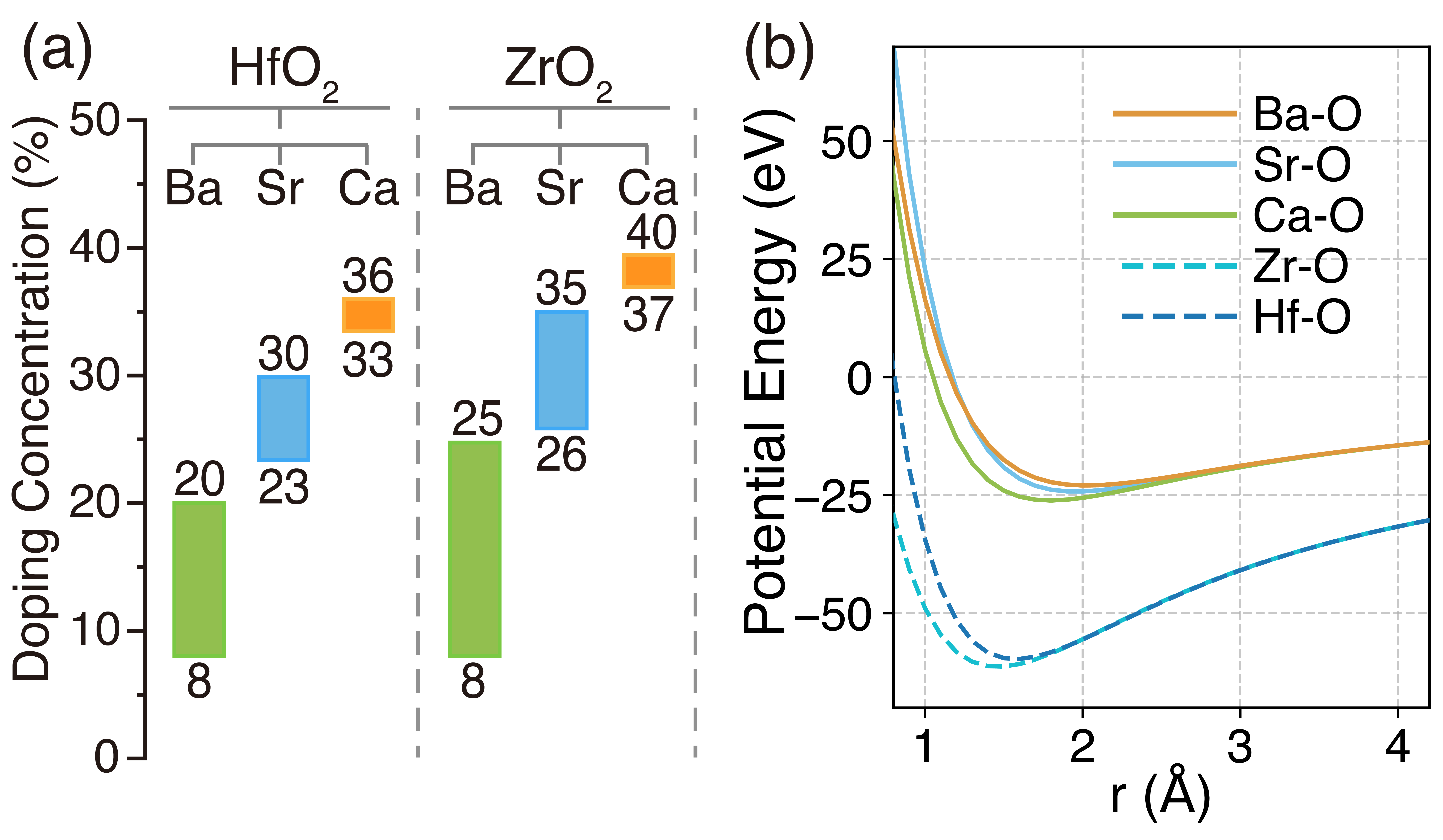}
\caption{
(a) Doping concentration ranges for the formation of SA structures across various doped systems. (b) Potential energy as a function of metal-oxygen dimer distance (core-core distances in core-shell models).
}
\label{fig:generality}
\end{figure}

We further demonstrate that the doping-controlled SA phase, with its superior properties, is of broad interest and general applicability. For instance, we dope HfO$_2$ with other alkaline earth metals from the same group as Ba (i.e., Sr and Ca) and reveal a similar concentration window for amorphization, as confirmed by XRD measurements (Fig.~S6) and summarized in Fig.~\ref{fig:generality}(a). These experimental findings are supported by MD simulations (Fig.~S7), where RDFs and BOP analyses yield results consistent with those presented in Figs.~\ref{fig:rdf} and \ref{fig:cages}, confirming the formation of SA structures in Sr- and Ca-doped HfO$_2$. Moreover, both experimental and computational evidence show that Ba, Sr, and Ca doping also induces the SA phase in ZrO$_2$ (Fig.~S8), suggesting that the SA structure is broadly achievable in fluorite-type systems.

Comparative analysis of the concentration windows across dopants reveals that the formation of the SA phase correlates with the ease of oxygen displacement from fluorite lattice sites. As shown in Fig.~\ref{fig:generality}(a), the critical concentration and width of the SA-forming window follow the trend Ba $>$ Sr $>$ Ca for both HfO$_2$ and ZrO$_2$. To rationalize this, we compute the interatomic potential between metal and oxygen atoms and find that Ba exhibits a flatter potential energy landscape than Sr or Ca (Fig.~\ref{fig:generality}(b)). This indicates that oxygen atoms in the vicinity of Ba experience a lower energy barrier for hopping from lattice to amorphous sites, thereby promoting greater disorder in the oxygen sublattice and facilitating SA phase formation.

\section{Conclusions}
In conclusion, we uncover a doping-induced polyamorphic transition in fluorite oxides, characterized by the emergence of a semi-ordered amorphous (SA) phase in Ba-doped HfO$_2$. This phase features an ordered metal sublattice coexisting with a disordered oxygen network. Different from conventional random-packing amorphous structures, the SA phase emerges from degenerate short-range symmetry breaking modes, which is consistent with Pauling's parsimony rule. Moreover, the SA structure exhibits enhanced thermodynamic stability, a wider bandgap, and a higher dielectric constant. These improvements originate from reduced subgap states and enhanced Born effective charges, driven by the semi-ordered local structure rather than direct dopant effects. The generality of this mechanism is confirmed in HfO$_2$ and ZrO$_2$ systems doped with Ba, Sr, and Ca, establishing the SA phase as a robust and broadly applicable structural motif. These findings provide fundamental insights into disorder engineering in complex oxides and offer a new strategy for designing high-performance gate dielectrics.

\section*{Acknowledgments}
Z. S. acknowledges support from the Double First-Class Initiative Fund of ShanghaiTech University (Grant No. SYLDX0342022) and computing resources provided by HPC platform of ShanghaiTech University. Z. W. acknowledges support from the Natural Science Foundation of China (Grant No. 52372113) and the Taishan Scholar Program of Shandong Province (Grant No. tstp20240511). The work of X. L. is supported by the National Natural Science Foundation of the People's Republic of China (Grants No. 11974211).

\bibliographystyle{apsrev4-2}
%\bibliography{manuscript-SA6.4.bib}
%apsrev4-2.bst 2019-01-14 (MD) hand-edited version of apsrev4-1.bst
%Control: key (0)
%Control: author (72) initials jnrlst
%Control: editor formatted (1) identically to author
%Control: production of article title (-1) disabled
%Control: page (0) single
%Control: year (1) truncated
%Control: production of eprint (0) enabled
%

\end{document}